\documentclass[10pt,twocolumn,letterpaper]{article}

\usepackage{cvm}
\usepackage{times}
\usepackage{epsfig}
\usepackage{graphicx}
\usepackage{amsmath}
\usepackage{amssymb}

\usepackage[pagebackref=true,breaklinks=true,letterpaper=true,colorlinks,bookmarks=false]{hyperref}

\cvmfinalcopy

\def\cvmPaperID{****} 

\ifcvmfinal\fi
\begin{document}

\title{DEMC: A Deep Dual-Encoder Network for Denoising Monte Carlo Rendering}

\author{Xin Yang\\
Dalian University of Technology\\
Dalian, P.R.China\\
{\tt\small xinyang@dlut.edu.cn}
\and
Wenbo Hu\\
The Chinese University of Hong Kong\\
Hong Kong S.A.R.\\
{\tt\small wbhu@cse.cuhk.edu.hk}
\and
Dawei Wang\\
The University of Hong Kong\\
Hong Kong S.A.R.\\
{\tt\small dawei@hku.hk}
\and
Lijing Zhao\\
Dalian University of Technology\\
Dalian, P.R.China\\
{\tt\small 785897146@mail.dlut.edu.cn}
\and
 Baocai Yin\\
Dalian University of Technology\\
Dalian, P.R.China\\
{\tt\small ybc@dlut.edu.cn}
\and
 Qiang Zhang\\
Dalian University of Technology\\
Dalian, P.R.China\\
{\tt\small zhangq@dlut.edu.cn}
\and
 Xiaopeng Wei\\
Dalian University of Technology\\
Dalian, P.R.China\\
{\tt\small xpwei@dlut.edu.cn}
\and
Hongbo Fu\\
City University of Hong Kong\\
Hong Kong S.A.R.\\
{\tt\small fuplus@gmail.com}
}

	\twocolumn[{%
		\renewcommand\twocolumn[1][]{#1}%
		\maketitle
		\begin{center}
			\centering
			\mbox{} \hfill
			\includegraphics[width=1.0\textwidth,height=5cm]{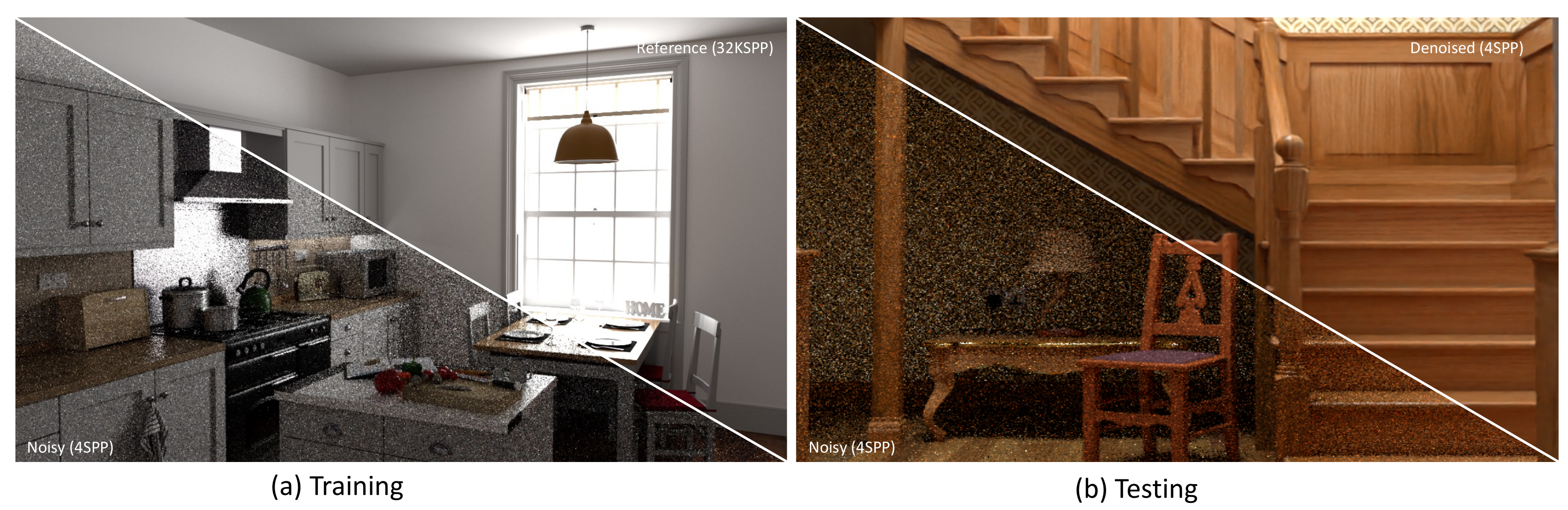}
			{\footnotesize{Figure 1.~} We propose a deep Dual-Encoder network for denoising Monte Carlo rendering to produce high quality images. We train our network to learn the complicated relationship between noisy images with low sampling rate and corresponding reference with high sampling rate (a). The learned model is then applied to denoise other rendering result with low sampling rate to predict noise-free results (b).}
			\hfill \mbox{}
		\end{center}%
	}] \par 

\begin{abstract} 
		In this paper, we present DEMC, a deep Dual-Encoder network to remove Monte Carlo noise efficiently while preserving details. Denoising Monte Carlo rendering is different from natural image denoising since inexpensive by-products (feature buffers) can be extracted in the rendering stage. Most of them are noise-free and can provide sufficient details for image reconstruction. However, these feature buffers also contain redundant information, which makes Monte Carlo denoising different from natural image denoising. Hence, the main challenge of this topic is how to extract useful information and reconstruct clean images. To address this problem, we propose a novel network structure, Dual-Encoder network with a feature fusion sub-network, to fuse feature buffers firstly, then encode the fused feature buffers and a noisy image simultaneously, and finally reconstruct a clean image by a decoder network. Compared with the state-of-the-art methods, our model is more robust on a wide range of scenes, and is able to generate satisfactory results in a significantly faster way. 
\end{abstract}  

\section{Introduction}

Producing a photo-realistic image from 3D models needs complex computation at every pixel of the image. For example, a ray tracing algorithm requires computing complex integral over all the ray paths between light source(s) and every point on image sensors. Monte Carlo (MC) raytracing~\cite{jensen2003monte} introduces a method to approximate this complex integral by tracing light path in a multi-dimensional space, in order to obtain an estimated value of the integral expression. Although Monte Carlo rendering has been widely accepted by many movie production studios, it suffers from noise pollution, which can only be mitigated by increasing the number of samples exponentially, making the synthesis of a noise-free and photo-realistic image very time consuming. However, some industry applications, such as real-time game rendering, virtual/augmented reality, require rendering high-quality images in a faster way.

Recently, a variety of methods~\cite{rousselle2013dfc,LBF,bitterli16nonlinearly,Bako17} for accelerating Monte Carlo rendering have been proposed. The core idea of these methods is to render a noisy image with a few samples per pixel (SPP) firstly, and then use denoising algorithms to reconstruct a perceptually noise-free image from the noisy image and auxiliary feature buffers. Here, the auxiliary feature buffers are inexpensive by-products generated in rendering stage, which contain geometry and texture information extracted from the 3D model. The auxiliary feature buffers are highly correlated with noisy images, and can conserve edge information. Most of them are noise-free and can provide sufficient details for image reconstruction. However, there is also redundant information mixed in the auxiliary feature buffers. This makes MC denoising different from natural image denoising. 
Hence, the main challenge of this problem is how to extract useful information that correlates with noisy RGB images from the auxiliary feature buffers to assist the reconstruction of clean images.
To address this problem, Moon et al.~\cite{moon2014adaptive} applied a linear model to approximate the ground truth, by weighting the error of each pixel based on the auxiliary features. Bitterli et al.~\cite{bitterli16nonlinearly} constructed collaborative regression using feature buffers. Kalantari et al.~\cite{LBF} built a network, using feature buffers, to predict parameters for cross-bilateral filter. Recently, Bako et al.~\cite{Bako17} proposed a deep convolutional network, leveraging feature buffers, to predict filter kernels for each individual pixel.

In this paper, we propose a deep network structure for denoising Monte Carlo rendering, Dual-Encoder network, to encode the feature buffers and noisy images with different encoders, and then use a decoder network to efficiently reconstruct clean images directly. Our proposed architecture includes two encoders: a feature buffer encoder for extracting the detail information in order to enhance image reconstruction in the decoding stage, and an HDR image encoder, which can transform the noisy image to a compact representation of the spatial contexts for the image. Since our feature buffers contain multiple channels, we introduce a feature fusion sub-network to merge feature buffers into three channels, which can extract edges and omit redundant information. Our network structure is shown in Figure~\ref{fig:network}. Compared with the state-of-the-art methods, our model is more robust on a wide range of scenes, and generates satisfactory results in a significantly faster way.

\section{Related Work}

After Cook ~\cite{cook1984distributed} published their paper ``Distribute Ray Tracing'', lots of researchers devoted to reconstructing the Monte Carlo rendering, and these works can be divided into two categories: 1) Traditional algorithms that rely on statistical analysis and process of sampled data in image-space or enhance MC renderings with information derived from an analytical analysis of the light transport equations. 2) Machine learning based methods that leverage machine learning algorithm to learn complex relationship between noisy images, feature buffers and references. 
\setcounter{figure}{1}
\begin{figure*}[tbp] 
	\centering
	\mbox{} \hfill
	\includegraphics[width=1\linewidth]{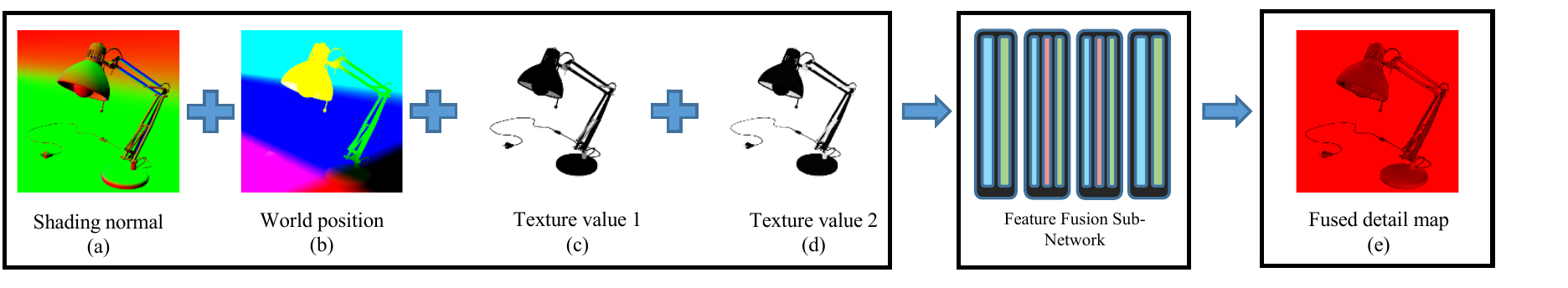}
	\hfill \mbox{}
	\caption{\label{fig:features}%
		Feature buffers extracted from the Tungsten renderer consist of shading normal (a), world position (b), texture values for the first and second intersections in RGB format for each sample during rendering (c, d), which are illustrated as RGB format. Then we feed them to our feature fusion sub-network to generate the fused detail map as shown in (e).
	}
\end{figure*} 
\subsection{Traditional Algorithms} 
In 2015, Zwicker et al.~\cite{zwicker2015recent} summarized non-machine learning algorithms, and divided them into two general classes: the priori methods and the posteriori methods. The priori methods leverage information acquired from an analysis of the light transport equations to enhance Monte Carlo samples and then generate adaptive reconstruction filters based on this information. For example, Ramamoorthi et al.~\cite{ramamoorthi2007first} apply derivative analysis to enhance adaptive sampling and conduct a more comprehensive and thorough first-order analysis of lighting, shading, and shadows in direct illumination. Jarosz et al.~\cite{jarosz2012theory} improved this work by applying a second-order analysis of indirect illumination. On the other side, the posteriori methods are used to leverage a family of reconstruction filters and develop errors estimation for reconstruction results. These approaches migrated from natural image denoising methods, and treated the renderer as a black box. For instance, Bauszat et al. ~\cite{bauszat2011guided} applied guided filter for removing MC rendering noise. Rousselle et al.~\cite{rousselle2013dfc} leveraged NL-mean filter for denoising. Moon et al.~\cite{moon2014adaptive} applied a linear model to approximate the ground truth. Bauszat et al.~\cite{bauszat2015general} constructed a robust error estimation method for MC rendering. Bitterli et al.~\cite{bitterli16nonlinearly} designed collaborative non-linear regression for reconstructing clean images. In summary, traditional methods generally need to select filter models or filter parameters manually, and require user intervention to empirically pick a suitable result. In comparison, our network will predict it automatically.

\subsection{Learning-Based Methods}
It is worth noting that Kalantari et al.~\cite{LBF} introduced a machine learning approach to the MC denoising field for the first time, though learning based methods have obtained great success on natural image denoising. They build a Multiple layer Perceptron (MLP) to predict parameters for cross-bilateral filter. Although it can avoid limitations caused by manually selecting parameters, it still inherits limits from a fixed filter (cross-bilateral filter or someone else). Recently, Bako et al.~\cite{Bako17} presented a Kernel-Predict Convolutional Network (KPCN), which uses deep convolutional network, dividing noisy image into two components and leveraging feature buffers as network inputs, to predict filter kernels for each individual pixel. KPCN completely solves the drawback from fixed filter, but the basis of this method is still a confirmed range filter kernel, its receptive field is highly limited. Yang et al.~\cite{FRCNN} presented a deep CNN for Monte Carlo rending reconstruction. They designed an end-to-end network, feed feature buffers and noisy image to the network directly. Although their work's receptive field could be unlimited, they didn't consider the diference between feature and RGB images, which makes their network hard to converge. Besides denoising MC rendering image, Chaitanya et al.~\cite{chaitanya2017interactive} proposed a recurrent autoencoder to reconstruct MC image sequence. Compared to denoising single image, there is temporal consistence priori could be employed in image sequence denoising. Therefore, denoising single MC rendering image is more challenging.

%
%
%

\subsection{Convolutional Network for Natural Image Denoising}
In addition to methods mentioned above, deep learning methods, especially deep convolutional neural networks (CNNs), have also shown great performance for natural image denoising problem. For example, Zhang et al.~\cite{zhang2017beyond} proposed a deep CNN for removing Gaussian noise, Gharbi et al.~\cite{gharbi2016deep} used CNN for demosaicking and denoising. Mao et al.~\cite{mao2016image} introduced a U-Net variant autoencoder to perform natural image restoration. Although these networks have obtained good performance in denoising problem, if we naively concatenate the auxiliary feature buffers with noisy MC rendering images and feed them into these image denoising networks, they cannot generate satisfactory results comparing to other MC denoising models. This is because the auxiliary feature buffers have different natures with RGB images, and without specifically designed structure for them the image denoising networks can not deal well with the auxiliary feature buffers.

\section{Methodology} %
In this section, we will introduce technical background and terminology briefly, and then describe the structure of our Dual-Encoder network.

\subsection{Problem Formulation} 
The goal of MC denoising is to predict noise-free images from noisy images and auxiliary features. For natural image denoising, noisy image is the only input. In contrast, we can get auxiliary features together with noisy images from the renderer for MC denoising. Specifically, as shown in Figure~\ref{fig:features}, the renderer output shading normals (i, j, and k), the world positions in Cartesian coordinates (x, y, and z) and texture values for the first and second intersections in RGB format for each sample as the auxiliary features (12 channels in total)~\cite{LBF}. Therefore, the per-pixel input $x_p = \{c_p,f_p\}$ is a vector of 15 channels, where $c_p$ is the tone-mapped color values and $f_p$ is the normalized auxiliary features. The details of tone mapping and normalization methods will be described in Sec. \ref{Training}. 

It is noteworthy that shading normals contain most of the geometry information of the scene, world positions have spatial location clue of the objects in the scene and texture values for the first and second intersections include the texture details information of the scene. They are all very helpful for reconstructing noise-free images, since geometry and spatial location clue are highly correlated with the structure edges in the images and texture details are corresponding to the texture edges in the images, besides, they are noise-free even rendered at very low SPP rate in most cases (except for using complicated camera models). Therefore, the main challenge is how to effectively explore the auxiliary features when reconstructing a clean image from the noisy image.

\begin{figure*}[!ht] 
	\centering
	\mbox{} \hfill
	\includegraphics[width=0.9\linewidth]{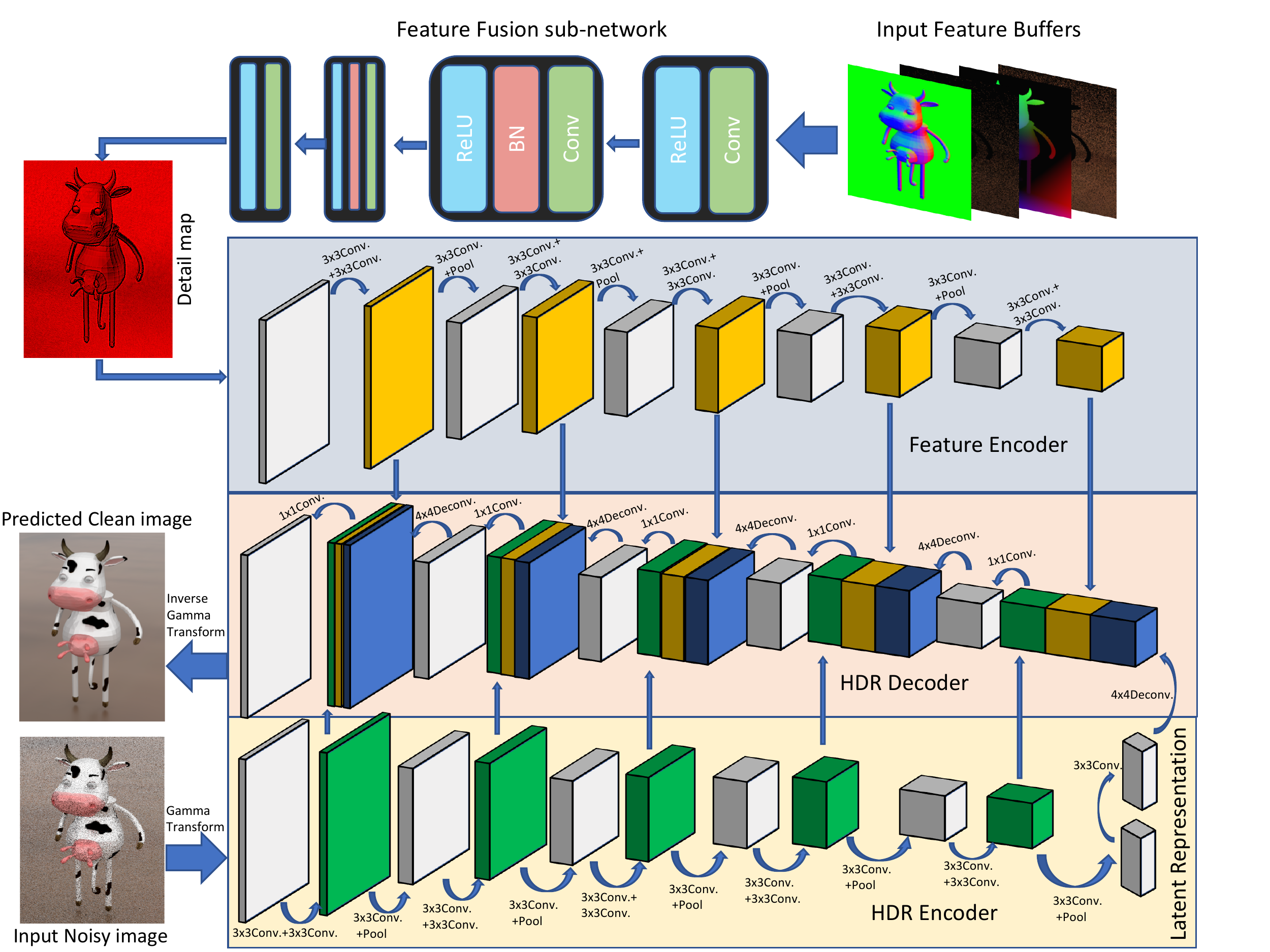}
	\hfill \mbox{}
	\caption{\label{fig:network}%
		A deep Dual-Encoder network, used for denoising Monte Carlo renderings. The feature buffers and noisy image are generated from renderer. The feature buffers are firstly fused by a feature fusion sub-network to get a detail map, and then the detail map and noisy image are encoded by the feature encoder and HDR encoder respectively. Finally, the latent representation is decoded to reconstruct a clean image with skip connection from the Dual-Encoder.
	}
\end{figure*} 


\subsection{Network Structure Design}
Denoising Monte Carlo rendering is different from natural image denoising since feature buffers as inexpensive by-products can be extracted in the rendering stage. Native networks for natural image denoising do not have a special structure to deal with feature buffers. Natural image denoising usually concentrates on color-based filter that exclude auxiliary buffer. Thus, native networks cannot work well on this problem. In Sec.\ref{sec:SEMC}, we tested Single Encoder Network (SEMC), which only has one encoder structure. As shown in Figure 4, our SEMC network structure contains as many trainable parameters as DEMC, which makes these two network structures have same trainable parameters totally. In this experiment, our DEMC's performance is better than SEMC, since SEMC cannot deal with auxiliary feature buffers well. Our goal is to predict the noise-free images from noisy images with the help of auxiliary features. The auto-encoder architecture~\cite{hinton2006reducing} 
can be used to transform data into a corresponding low-dimensional latent 
representation, and then reconstruct the original dimension data. Considering the characteristics of removing Monte Carlo noise with the rich information in the feature buffers, we extend the standard auto-encoder network with skip connection. We design a Dual-Encoder network for encoding feature buffers and noisy 
images simultaneously, and reconstructing corresponding clean images. The 
problem could be formulated as: 
\begin{equation}\label{goal} 
\hat{\theta}=\arg \mathop{\min}\limits_{\theta}\mathcal{L}(\overline{c}_p,\mathcal{G}(x_p,\theta)),
\end{equation} 
where $\mathcal{G}(x_p,{\theta})$ is our DEMC model, $\theta$ is the trainable parameters, and $\mathcal{L}$ is a loss function between the reference value $\overline{c}_p$, which is rendered with extremely high SPP (e.g., 32K) and predicted value $\mathcal{G}(x_p,\theta)$. We propose a Feature Fusion sub-network to deal with feature buffers. It is worth noting that edges in different areas of the image could be drawn from different features. Then, our Feature Fusion sub-network can merge the edges. As shown in Figure~\ref{fig:network}, compared to the reference, the feature fusion result contains more details and edges information. This sub-network contains four convolutional blocks. The input and output blocks both contain a convolutional layer, and a rectified linear unit (ReLU) activation layer, since ReLU can promote performance in multiple tasks and boosting the model convergence to the local minimum~\cite{balduzzi2016neural}. For the blocks in the middle, we add a batch normalization layer inside it for a better optimization~\cite{he2016deep}. The output of feature fusion sub-network contains three channels, which can preserve more detail and edge information, comparing to the original feature buffers, as shown in Figure~\ref{fig:features}. Since we train this feature fusion sub-network with our Dual-Encoder network jointly, the sub-network will automatically extract structure and texture details, which can be used for the reconstruction stage from feature buffers.

In the encoder network, there are three convolutional layers followed by a 
max-pooling layer. Each of these four layers constitutes a down-sampling unit. 
We employ two encoders to extract the information from RGB values and feature 
values separately, since the information representations of noisy images and 
feature buffers are different. Each encoder contains five down-sampling units, encoding the input images and feature buffers respectively in a ${W}/{32}  \times {H}/{32} \times 512 $ latent representation, where $W$ and $H$ indicate the width and height of the input data. We can obtain two latent representations through 
the Dual-Encoder network, with one corresponding to noisy image, and the other 
corresponding to the low-dimensional representation of the features. The 
representation of the noisy images will be used to feed into decoder network and reconstruct the final result since the information of the features can be 
transferred by the skip-connection structure. 

In the decoder network, we use deconvolutional layers with $4\times4$ kernel 
to up-sample images into $2W \times 2W $ scales. After five up-sampling 
layers, we can get the final results with the same resolution as the noisy images. 
All the convolutional layers and deconvolutional layers in our Dual-Encoder 
network use ReLU activation functions, which can promote network performance in multiple tasks and boosting the model convergence to the local minimum~\cite{balduzzi2016neural}. We use skip-connection to transfer 
each level of two encoders to the corresponding level of the decoder side 
simultaneously, since some information may be lost during the encoding stage 
and this information can be used by skip-connection to enhance the decoding 
stage. Skip-connection will concatenate the outputs of the layers from the 
two encoders and the corresponding decoder layer along the third dimension. 
To be more specific, given the outputs of two layers, whose dimensions are both 
$W \times H \times K$, the concatenated result will be $W \times H \times 
3K$. Then we use a convolutional layer with kernel size of $1 \times 1$ 
to fuse the concatenated layers into the output with $K$ channels. The 
skip-connection is defined as 
\begin{equation}\label{skip} \tilde{h^D_i} = \sigma (W\begin{bmatrix} 
h^D_i\\h^{Ef}_i \\ h^{Eh}_i \end{bmatrix}+b). \end{equation} The vector 
$h^D_i$, $h^{Ef}_i$ and $h^{Eh}_i$ denote the $i$th layer tensors from the 
corresponding decoder layer, feature encoder layer and HDR encoder layer, respectively. 
Meanwhile, $\tilde{h^D_i}$ is the decoder feature fused from skip-connection. 
$W$ is a weight matrix, $b$ denotes the bias of feature fusion, and $\sigma$ is the ReLU 
activation function. For instance, $h^D_i$, $h^{Ef}_i$ and $h^{Ei}_i$ are $k 
\times 1$ vectors and $W$ will be a $K \times 3K$ weight matrix, fusing three 
vectors into $K$ dimensions. For this purpose, we set the weights as 

\begin{equation}
\label{Wmatrix} W =\begin{bmatrix} I& I& I \end{bmatrix}, 
\end{equation}
where $I$ is the $k$th identity matrix and $b$ is set to 0.

\begin{center}
	\includegraphics[width=0.9\linewidth]{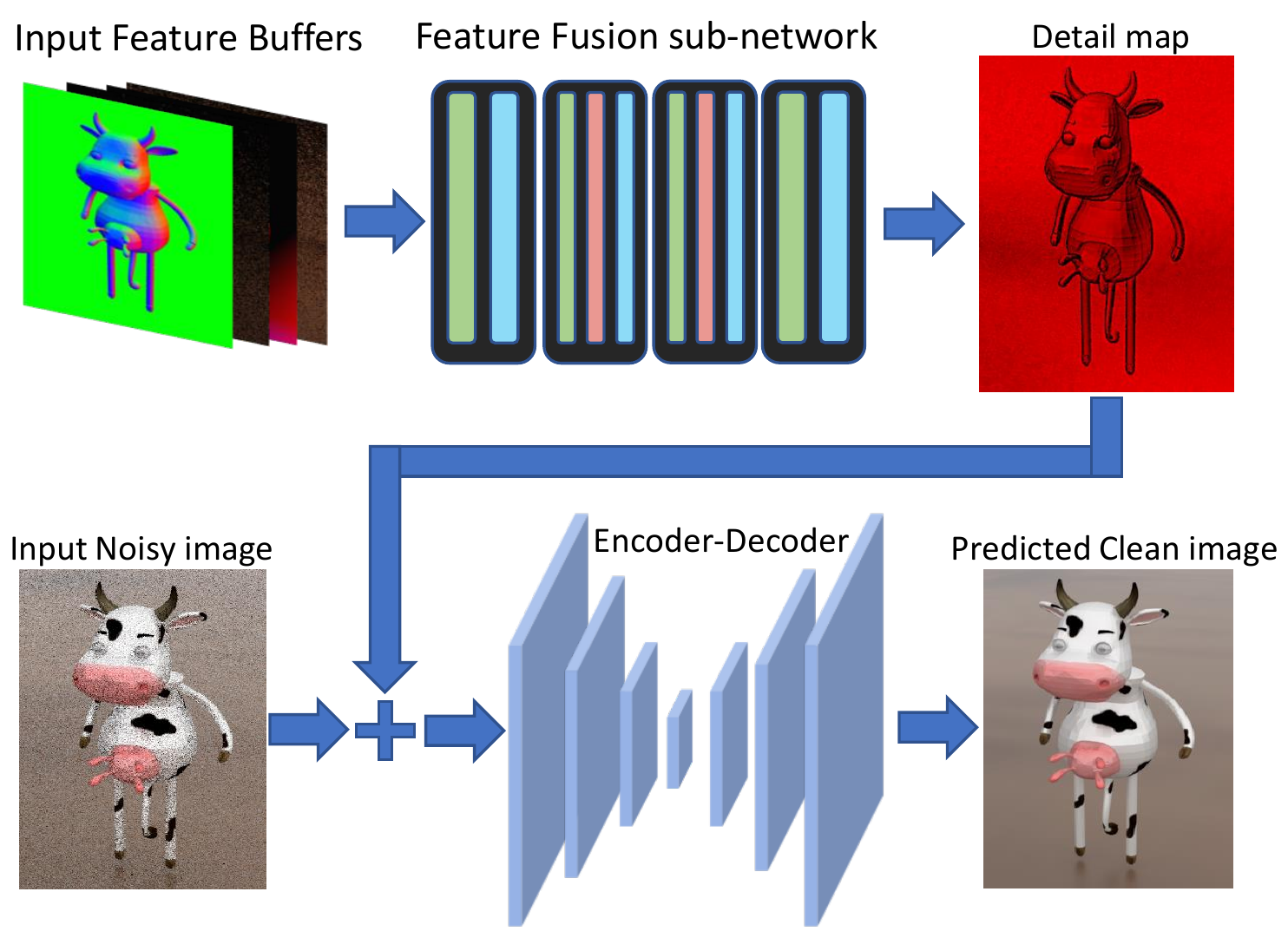}\\
	\vspace{2mm}
	\label{fig:SEMC_Net}%
	\parbox[c]{8.3cm}{\footnotesize{Figure 4.~} Network structure of Single Encoder Network (SEMC), which only has one encoder structure and contains as many trainable parameters as our Dual-Encoder Network (DEMC). }
\end{center}

\section{Experiments and Results} 

\subsection{Data}
A sufficiently large and effective dataset is needed to train a robust model for denoising MC renderings in a variety of distributed effects such as depth of field, area lighting, glossy reflections, and global illumination. We pick up some scenes from \emph{Blender-Swap} and~\cite{resources16}, clean up the geometry, then manually set up PBR materials, lighting and camera, and finally make them available for the Tungsten renderer. The ground truth images are rendered at 32k or higher SPP rate for production-level quality, while input noisy images are rendered at a fixed 4 SPP. For the training set, we select 97 scenes that cover different distributed effects to expand the generalization capability of our model. We cannot use KPCN's training set to train our model, since the training dataset of KPCN is not public. Meanwhile, our test set contains 36 scenes, which can represent different scene types. Our testing dataset and training dataset contain no similar images. Example images of the training set are shown in Figure 5.


\vspace{2mm}

\begin{center}
	\includegraphics[width=0.8\linewidth]{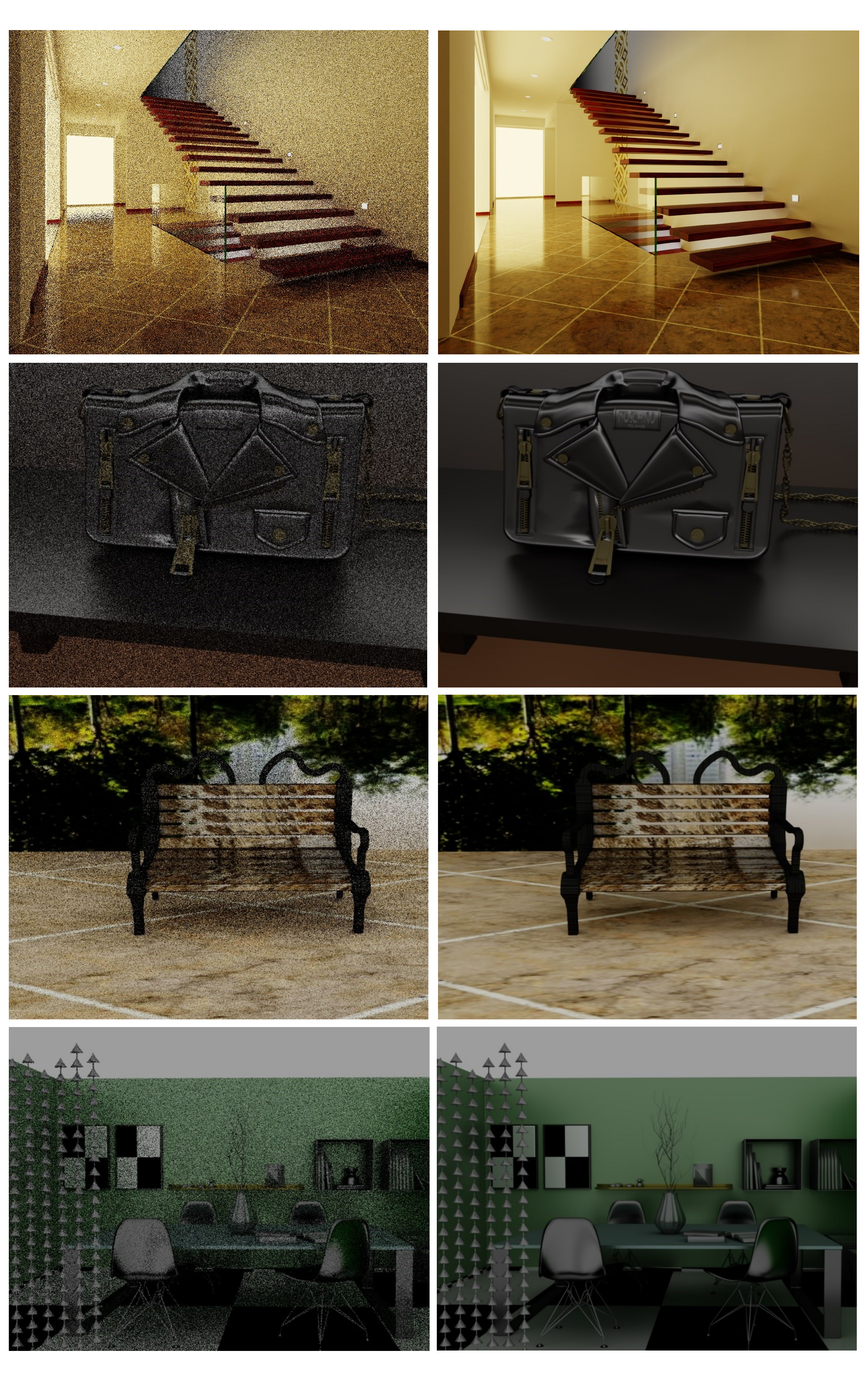}\label{fig:training set}
	\vspace{2mm}
	\parbox[c]{8.3cm}{\footnotesize{Figure 5.~}  Examples of the training set. The left column is input noisy images rendered with low SPP (e.g. 4 SPP), the right column is referenced images rendered with high SPP (32K SPP or higher). }
\end{center}

\vspace{1mm}


\vspace{2mm}

\begin{center}
	\includegraphics[width=0.8\linewidth]{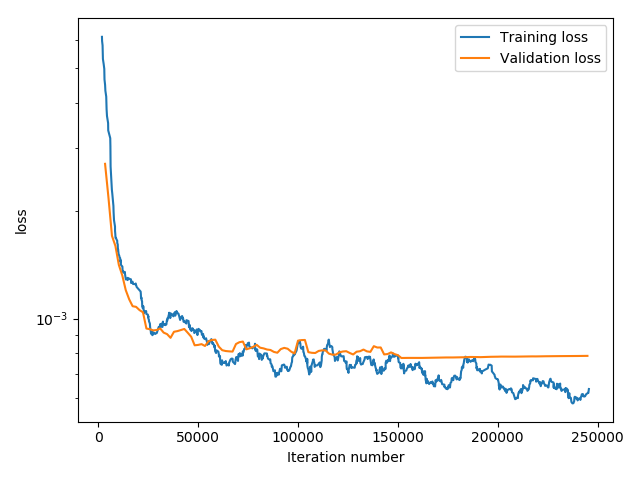}\label{fig:loss}
	\vspace{2mm}
	\parbox[c]{8.3cm}{\footnotesize{Figure 6.~}  We plot the training and validation loss against the number of iterations during the training stage. The data has been smoothed and is plotted in log domain for better visualization. }
\end{center}

\vspace{1mm}

\setcounter{figure}{6}
\begin{figure*}[htbp] 
	\centering
	\mbox{} \hfill
	\includegraphics[width=1\linewidth]{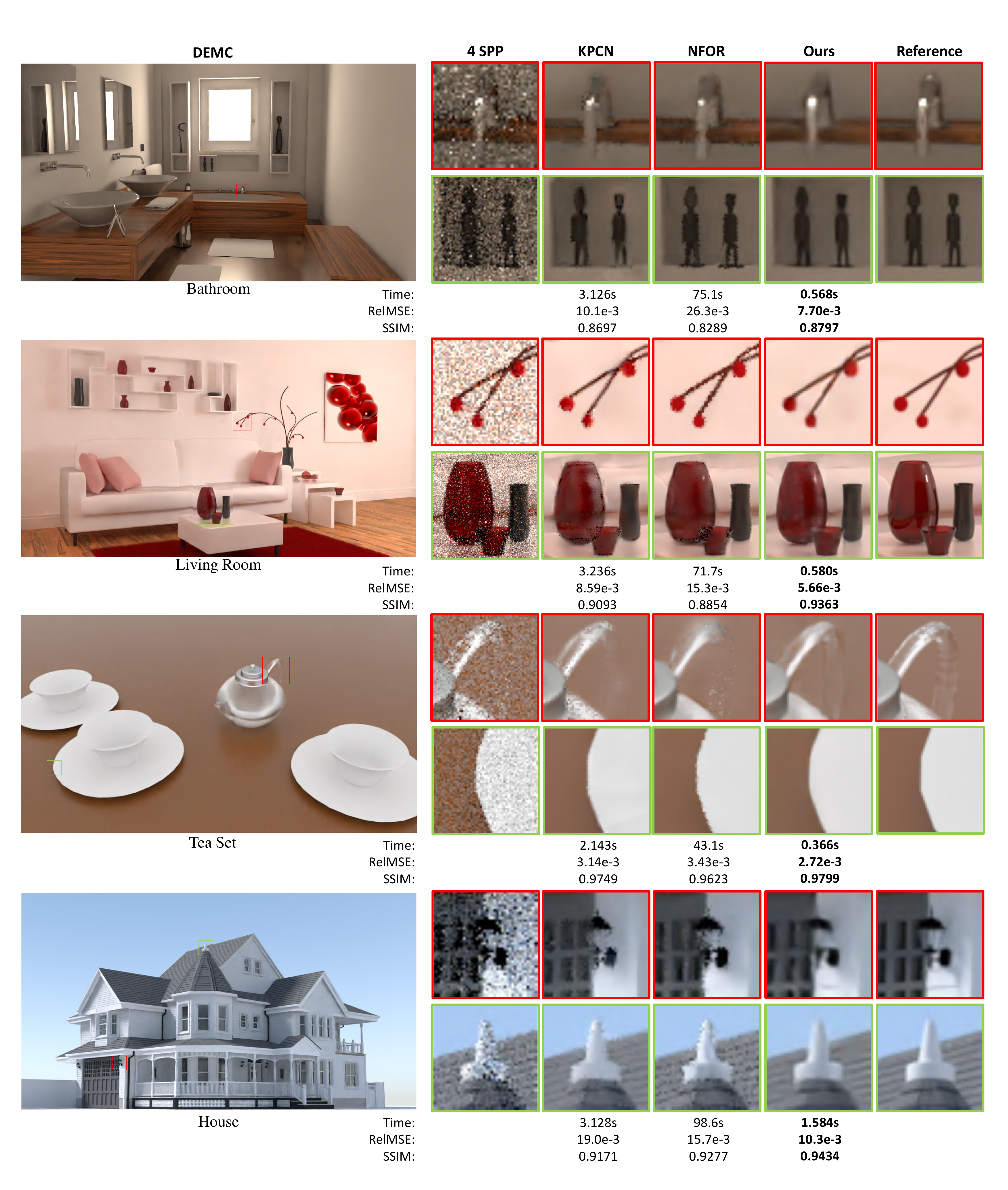}
	\hfill \mbox{}
	\caption{\label{fig:compare}%
		We compare our DEMC against the state-of-the-art methods, NFOR~\protect\cite{bitterli16nonlinearly} and KPCN~\protect\cite{Bako17}. Our method is faster than all the other methods while preserving more details. The RelMSE (relative MSE) and SSIM~\protect\cite{wang2004image} index values are listed below each close-up result (for RelMSE lower is better, for SSIM higher is better). Tone mapping of the insets has been adjusted equally for all algorithms for best visibility. The full results in the test set can be found in the supplementary materials.
	}
\end{figure*}

\begin{figure*}[!ht] 
	\centering
	\mbox{} \hfill
	\includegraphics[width=1\linewidth]{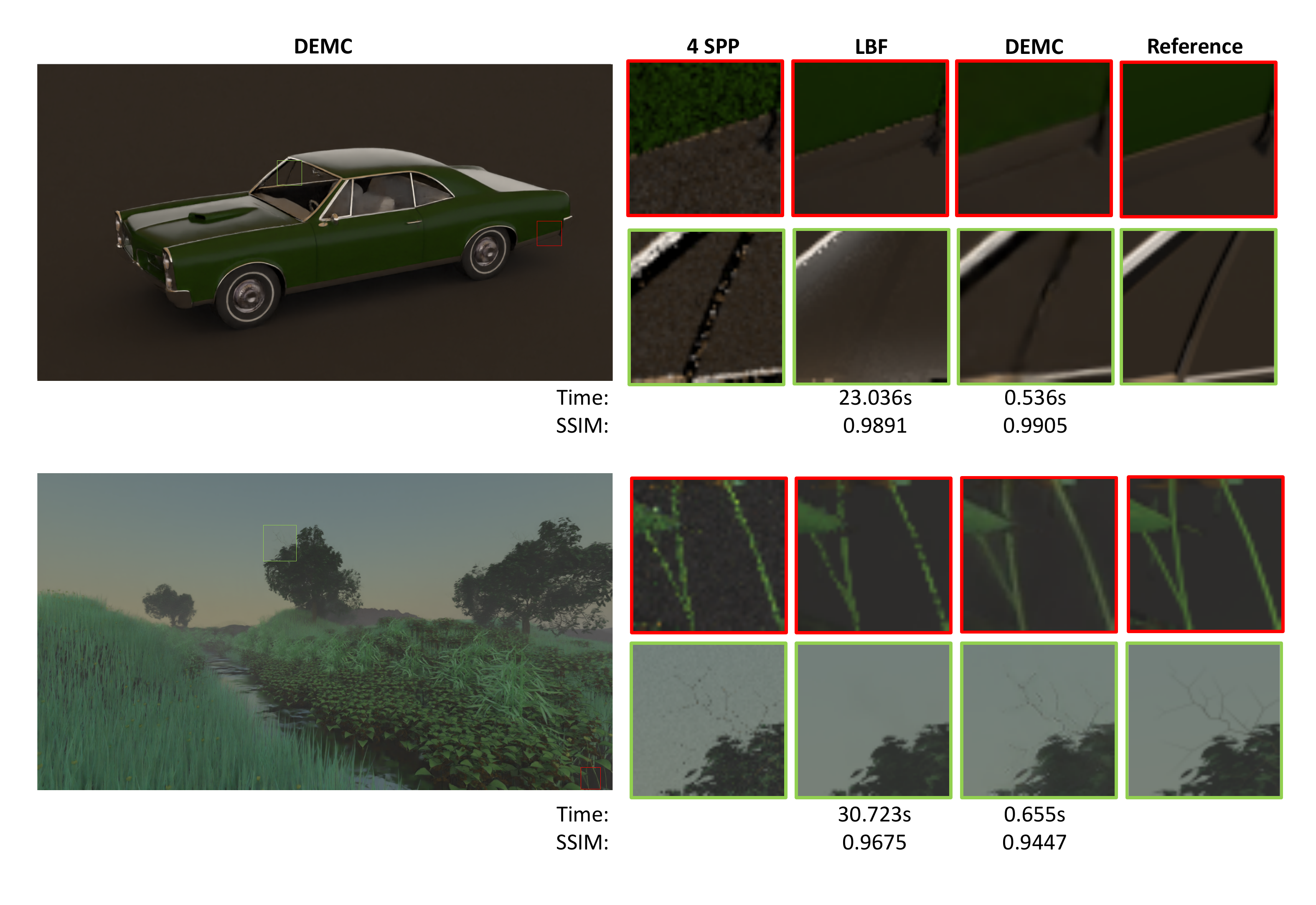}
	\hfill \mbox{}
	\caption{\label{fig:LBF}
		We compare our DEMC against LBF~\protect\cite{LBF} on the PBRT2 renderer~\protect\cite{pharr2016physically} to show our DEMC has high flexibility with different rendering systems. 
	}
\end{figure*}

\subsection{Training}\label{Training} 

Different from LDR (Low Dynamic Range) images, the noisy input HDR (High Dynamic Range) images have a large pixel value range. This make the training extremely unstable. Hence, we employ Gamma transformation on HDR images to compress pixel values. Similar technique has been used for training a neural network to inverse LDR image to HDR domain~\cite{eilertsen2017hdr}. The concrete transforming equation for the noisy HDR image $c_p^*$  is: 
\begin{equation}\label{gamma_transfer} 
c_p = (c_p^*)^{1/\gamma},
\end{equation} 
where $\gamma$ is set to 2.2 in our experiments. Similarly, auxiliary features also have a large value range. For instance, world position values are always large while shading normal values are small. But we do not need them in their original domain, we normalize them using common Z-score method.

Since the input noisy images are compressed with the Gamma transformation while the ground truth images are still in the HDR domain, we apply the inverse of Eq.~\ref{gamma_transfer} to the output of our network to transfer the predicted images back to the HDR domain. Then, we compute the loss between the constructed images and ground truth as follows: 
\begin{equation}\label{loss} \mathcal{L} = \frac{1}{N}\sum_{q \in \{r,g,b\}}\sum_{p=1}^{N}
\frac{(\overline{c}_{p,q} - \hat{c}_{p,q})^2 }{{\overline{c}^2_{p,q}} + 
	\epsilon}, 
\end{equation} 
where $N$ is the total number of pixels, $\hat{c}_{p,q}$ and $\overline{c}_{p,q}$ are the $q$th color channel of the reconstructed and ground truth pixels, respectively, and $\epsilon$ is a small number (0.001 in our implementation) to avoid division by zero. This metric is RelMSE~\cite{rousselle2011adaptive}, which can give higher weights to the regions where the ground truth images is darker, since the human visual system is more sensitive to color variations in darker regions~\cite{LBF}. We minimize the loss function in the HDR domain directly. By doing so, we can train the model to converge to the final optimal solution in HDR domain.

\begin{figure*}[!ht]
	\centering
	\mbox{} \hfill
	\includegraphics[width=1\linewidth]{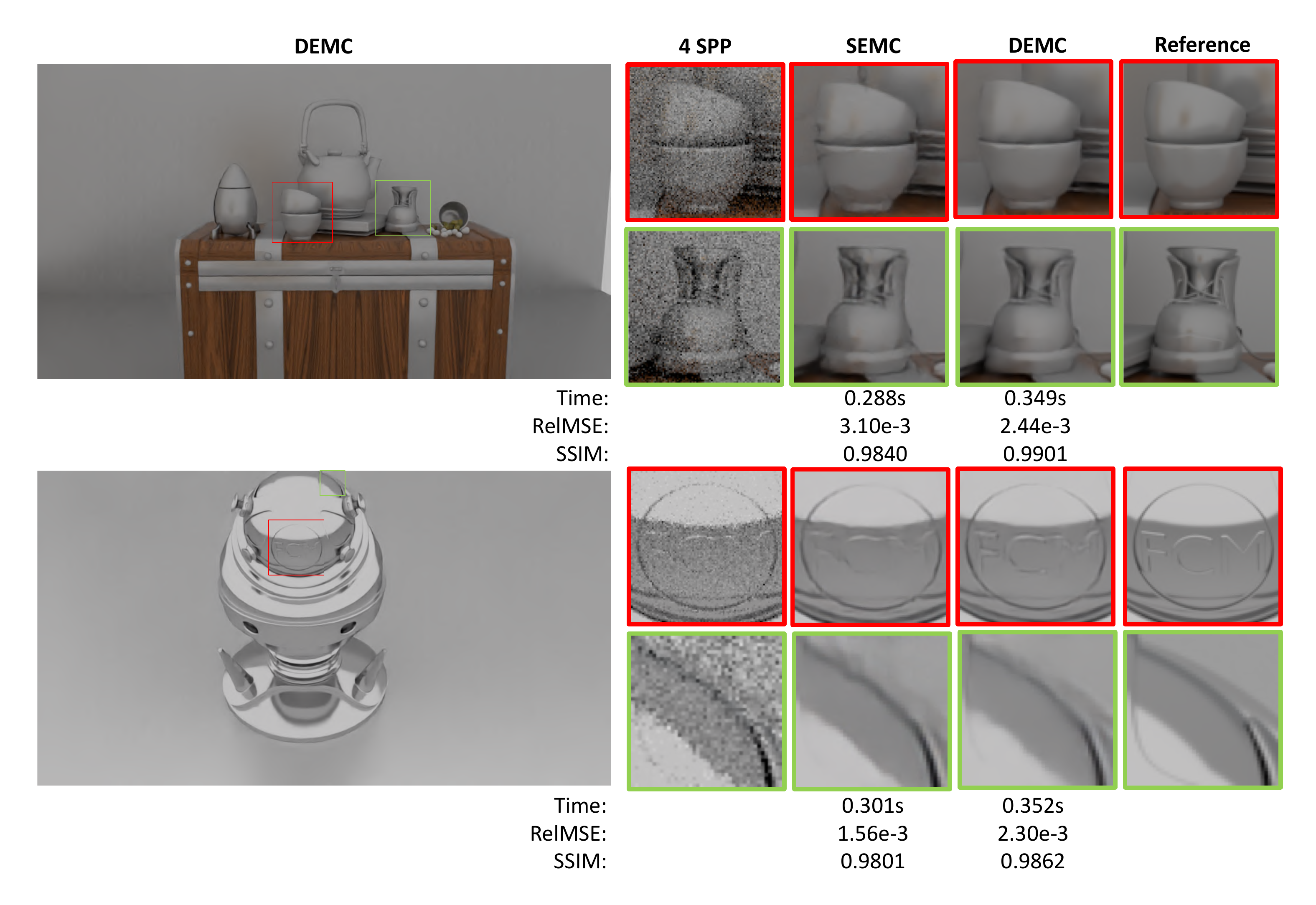}
	\hfill \mbox{}
	\caption{\label{fig:SEMC}%
		We compare our DEMC against Single Encoder Network (SEMC) to show the strong representation capability of our DEMC. They are all executed on noisy images and auxiliary features rendered at 4 SPP. Tonemapping of the insets has been adjusted equally for all the algorithms for best visibility. 
	}
\end{figure*}

We initialize the weights of our DEMC network with different strategies for different parts. For the convolutional layers of the encoder network and latent representation, we use the Xavier method~\cite{glorot2010understanding} to initialize them. For the deconvolutions of the decoder network, we initialize them using a bilinear up-sampling matrix. The skip-connections are initialized as Eq.~\ref{Wmatrix}. 

We implemented our DEMC using Tensorflow~\cite{abadi2016tensorflow} on Ubuntu with GPU acceleration. We set patch size as $128 \times 128$ and stride as $80$. Finally, we get about 57k patches for training. In the training stage, our DEMC is optimized with ADAM~\cite{Kingma2014Adam}, and the learning rate range is set from $10^{-4}$ to $5 \times 10^{-6}$. Our experiments are executed on a PC with Intel Core i7 7700k, NVIDIA GTX 1080Ti and 32G memory. The network is trained for approximately 250K iterations over the course of about 2 days. The training and validation loss log are shown as Figure 6.

\subsection{Comparison Against State-of-the-art} 
We compare our proposed method, DEMC, against the state-of-the-art approaches, NFOR~\cite{bitterli16nonlinearly}, KPCN~\cite{Bako17} and LBF~\cite{LBF}. For NFOR, we use the authors' open source implementation, which is plugged into the Tungsten renderer. For KPCN, we use the model trained by the author for Tungsten renderer. For LBF, we use the authors' original implement on the well-known renderer, PBRT2~\cite{pharr2016physically}. We retrain our model on the scenes of PBRT2 to compare our model against LBF on some test scenes that is not intersected with the training set. The reason why we do not retrain LBF on Tungsten renderer scenes is that auxiliary feature buffers extracted from Tungsten and PBRT2 are slightly different, especially for texture values for the second intersection and visibility, and the LBF model we trained on Tungsten scenes does not performs as well as the original one trained on PBRT2 scenes. Therefore, we compare our model against NFOR and KPCN in Tungsten renderer while against LBF in PBRT2 renderer.

In this paper we focus on the applications that require rendering high-quality images in a fast way, e.g., game rendering, virtual/augmented reality and prototype design as stated in the introduction. For such applications, rendering speed is very important, while high samples per pixel (SPP), e.g. 32 SPP, will greatly slow down the rendering.
Therefore, all these methods are tested with noisy images and feature buffers rendered at 4 SPP, while reference images are rendered at 32K SPP or higher to make sure they are perceptually noise-free. 
To assess the performance of these methods, we use 2 metrics, RelMSE (Relative MSE) and SSIM (Structural Similarity Index) whose values are from 0 to 1, where 1 indicates perfect quality with respect to the ground truth image. The reported time shown in Figure~\ref{fig:compare} means the denoising time, exclude the time for rendering the noisy images, since all of these methods take the similar time to generate them.

In Figure~\ref{fig:compare}, we show a subset of the comparison results in our test set, and our DEMC performs better than the state-of-the-art methods both perceptually and quantitatively. The full results in the test set can be found in the supplementary materials. For instance, in the \emph{Bathroom} scene, our method could reconstruct both object structure and high-light reflection due to the helpful auxiliary features and the strong representation capability of our DEMC, while KPCN generates residual noise in high-light regions and NFOR suffers from noisy edges in complex geometry regions. In the \emph{Tea Set} scene, KPCN and NFOR are both blurred in teapot handle region, which contains the refraction and reflection of rays. In contrast, our method can represent the teapot handle region more accurately compared to the reference. In the \emph{House} scene with global illumination, compared with KPCN and NFOR, our method could reconstruct a cleaner result while preserving more details. In terms of speed, for an image of $1280\times720$, our DEMC takes about 0.6 seconds to evaluate and output a fully denoised image, while the GPU based method, KPCN, needs more than 3.0 seconds for the same image\protect\footnotemark[1]. 

\protect\footnotetext[1]{Note that, the noisy image and auxiliary features 
	rendered time is not included here, since all of the methods takes the same 
	time to generate them approximately.} 


\vspace{2mm}

\begin{center}
	\includegraphics[width=1\linewidth]{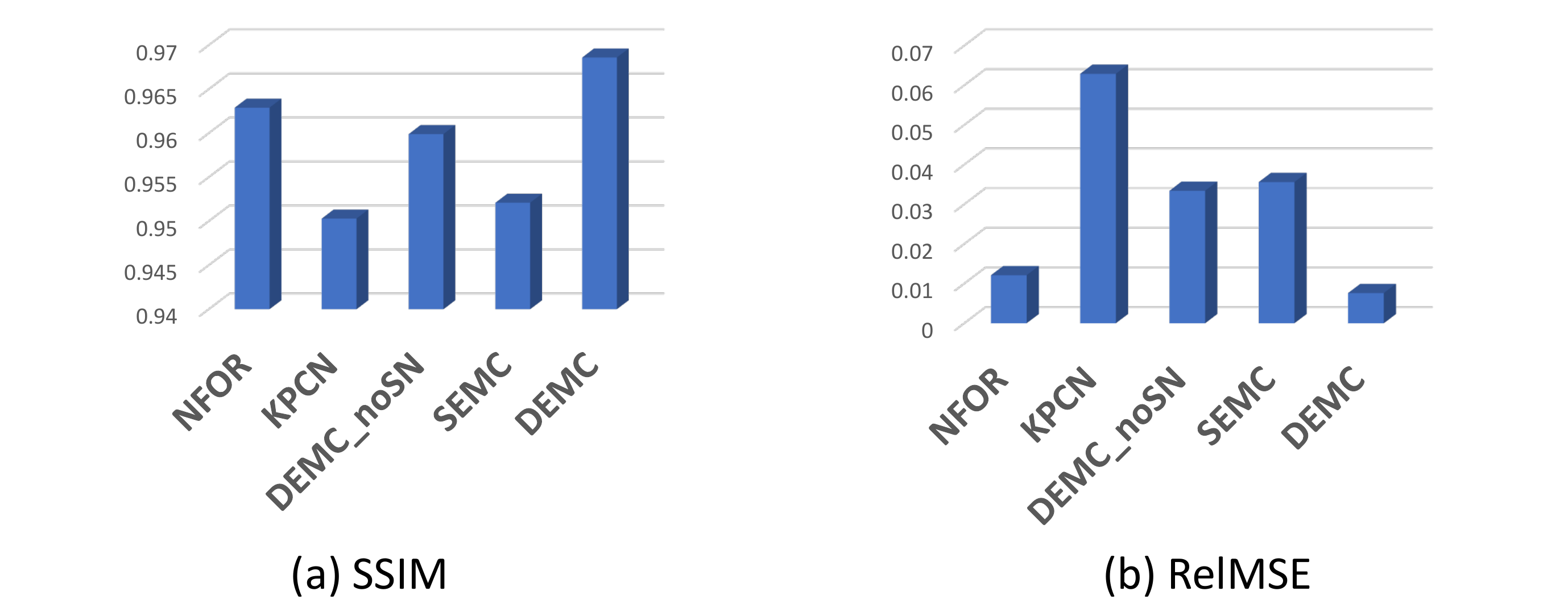}\label{fig:metric}
	\vspace{2mm}
	\parbox[c]{8.3cm}{\footnotesize{Figure 10.~}  Average performance of NFOR~\cite{bitterli16nonlinearly}, KPCN~\cite{Bako17}, Single Encoder Network (SEMC) and our DEMC model across our test scenes on 4spp. In order to demonstrate that feature fusion sub-network structure is benefit to the final performance, we also test our DEMC model without feature fusion sub-network structure (DEMCnoSN). (a) shows the performance in the metric of SSIM, higher SSIM meaning better performance. (b) shows the performance in the metric of relative MSE, which is also our loss function, shown in Equation~\ref{loss}, lower RelMSE meaning better performance. }
\end{center}

\vspace{1mm}

Figure 10 shows a comparison of the average performance of our DEMC model and KPCN~\cite{Bako17} and NFOR~\cite{bitterli16nonlinearly}, on the Tungsten renderer. Our DEMC model outperforms the other two methods, in both the error metrics, namely SSIM and relative MSE. In order to demonstrate the effectiveness of feature fusion sub-network, we evaluated our DEMC model without it (DEMCnoSN in 10). For DEMCnoSN, we directly feed the auxiliary feature buffers to the feature encoder and noisy images to another encoder and reconstruct the clean images as DEMC do. As shown in Figure 10, the feature fusion sub-network structure obviously promoted the performance of DEMC model. We did not include PSNR as one of the error metrics, since there is a parameter in PSNR's definition, which indicates the maximum value of images. But as we mentioned above, the values in HDR images can be positive infinity theoretically. Therefore, the PSNR metric is not suitable for evaluating HDR images. 

We retrained our DEMC on a training set of PBRT2 scenes, including 50 different scenes, to compare with LBF. The example results in test set is shown in Figure~\ref{fig:LBF}, wich contain scenes with both high frequency features such as vegetation and low frequency features such as the surface of the car. We can see that, the results generated from LBF will over-smooth the scene, therefore some slight object structures will be erased while our results are more accurate and realistic. Our DEMC network outperforms LBF in SSIM metric, and generates result much faster than LBF.

\subsection{Dual-Encoder Network \emph{vs.} Single Encoder Network}\label{sec:SEMC}
We conduct an experiment to show our Dual-Encoder network's performance, 
comparing against single encoder. We design a Single-Encoder network (named 
SEMC for convenience), which has the same feature fusion sub-network and the 
same input as DEMC. To be more specific, the auxiliary features are fused 
by the feature fusion sub-network firstly, then the fused detail map is 
concatenated together with the noisy image, and finally the concatenated data is flowed into a standard auto-encoder and decoder network with skip connection. We train the SEMC using the same training set and hyper parameters as DEMC. As shown in Figure~\ref{fig:SEMC}, we show the qualitative and quantitative comparison results of DEMC and SEMC for \emph{Silver-Material} and \emph{Low-Design} scenes. Compared to SEMC, DEMC can preserve more structure and shadow details. In Figure 10, we show quantitative comparison between DEMC and SEMC on our test set, which shows that the DEMC performs better than SEMC on the SSIM metrics. This shows that the Dual-Encoder network structure can more effectively extract useful information that correlates with noisy RGB images from the auxiliary feature buffers to assist the reconstruction of clean images than Single-Encoder network structure in MC denoising.

\section{Limitations and Future Work}

Our method belongs to auxiliary feature buffers based methods, which is the same as LBF~\cite{LBF}, KPCN~\cite{Bako17} and NFOR~\cite{bitterli16nonlinearly}. This kind of methods assume the features are highly correlated to noisy image, but such assumption is not always correct, such as the strong specular reflection scenes where the auxiliary features are less relevant to the noisy image. As shown in Figure 11, there are two glass balls on the ground, and the ball reflects texture and pattern of the ground. However, this detail is not shown in the feature \emph{texture1} \emph{world position} and \emph{shading normal}. Since this kind of scenes are very different from the other scenes, our model cannot deal with them well and the specular region will be blurred, which is also common among the auxiliary feature buffers based methods~\cite{Boughida2017Bayesian}. 

\vspace{2mm}

\begin{center}
	\includegraphics[width=0.8\linewidth]{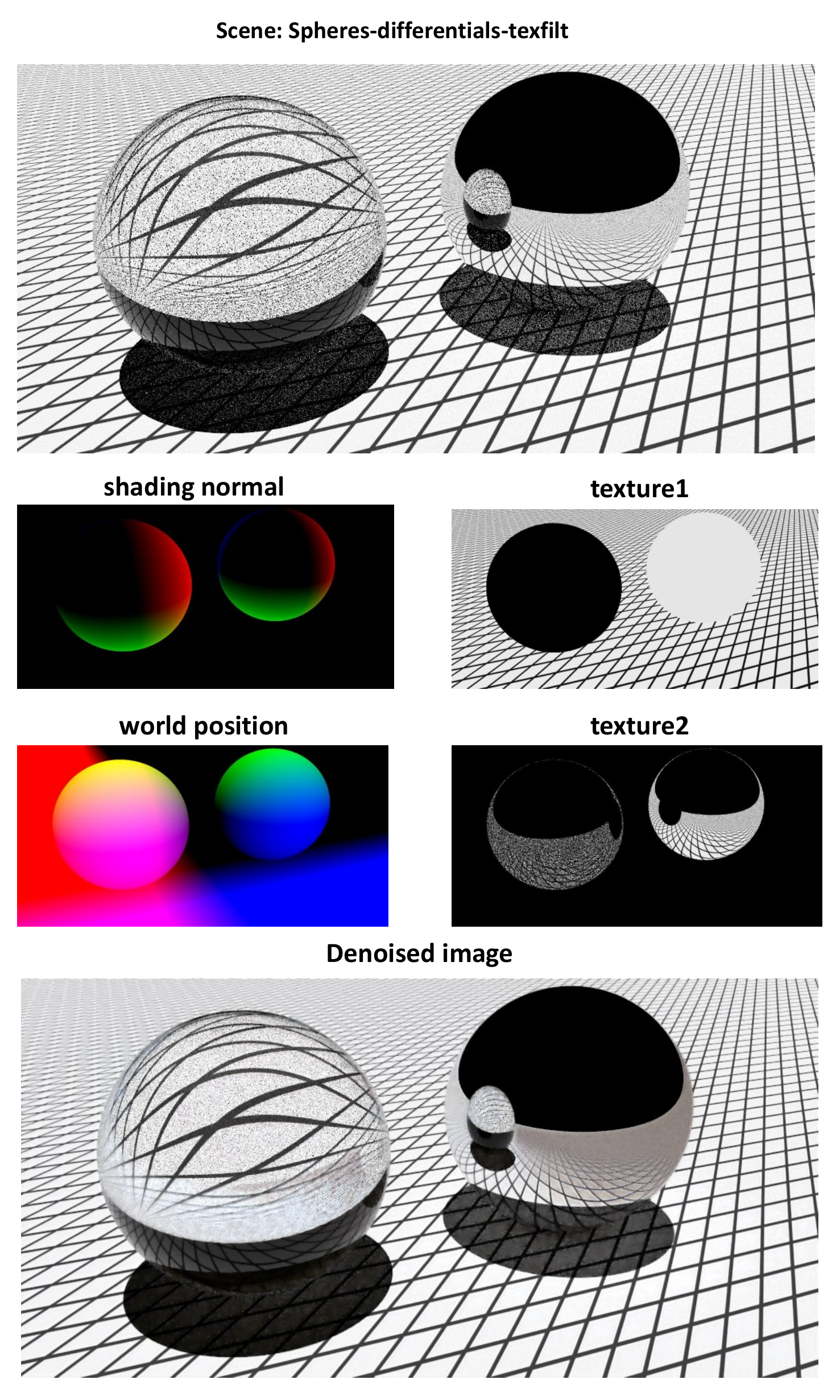}\label{fig:limitation}
	\vspace{2mm}
	\parbox[c]{8.3cm}{\footnotesize{Figure 11.~}  We visualized a scene named \emph{spheres-differentials-texfilt}, and its corresponding feature buffers. Since the material of balls are glass, there are lots of specular reflection effect. Our model cannot work well on this kind of scenes. }
\end{center}

\vspace{1mm}

Since our method is designed for single MC rendered image denoising, if directly applying it to an animated sequence in a frame-by-frame manner, the results may not be temporal coherent. A practical solution is using a video temporal consistency filter, such as Lang et al.~\cite{lang2012practical}, Bonneel et al.~\cite{bonneel2015blind}, Lai et al.~\cite{lai2018learning} and so on, to post process the denoised frames to get temporal coherent results. However, a better solution may be taking into account temporal coherence in the neural network by adding recurrent connections between previous frames and current one. We leave the investigation of such methods for future work.


\section{Conclusions}

In this paper, we have presented a novel Dual-Encoder network (DEMC) for denoising Monte Carlo renderings. We also proposed a feature fusion sub-network, which can be trained jointly with the Dual-Encoder network to extract structure and texture details from auxiliary features. Benefited from the strong representation capacity of the Dual-Encoder and feature fusion sub-network, our method can effectively explore the auxiliary features to help denoise MC renderings. In contrast to the state-of-the-art MC denoising approaches that based on either machine learning or not, our model is capable of reconstructing MC renderings both effectively and efficiently. 

\section{Acknowledgement}
This work was supported in part by the National Natural Science Foundation of China under Grant 91748104, Grant U1811463, Grant 61632006, Grant 61425002, Grant 61772100, and Grant 61751203, in part by the National Key Research and Development Program of China under Grant 2018YFC0910506, in part by the Open Project Program of the State Key Lab of CAD\&CG (Grant No. A1901), Zhejiang University, the Open Research Fund of Beijing Key Laboratory of Big Data Technology for Food Safety under Project BTBD-2018KF, and in part by the Innovation Foundation of Science and Technology of Dalian under Grant 2018J11CY010.

	\bibliographystyle{cvm}
	\bibliography{cvmbib}

\begin{thebibliography}{10}\itemsep=-1pt

\bibitem{abadi2016tensorflow}
M.~Abadi, P.~Barham, J.~Chen, Z.~Chen, A.~Davis, J.~Dean, M.~Devin,
  S.~Ghemawat, G.~Irving, M.~Isard, et~al.
\newblock Tensorflow: A system for large-scale machine learning.
\newblock In {\em OSDI}, volume~16, pages 265--283, 2016.

\bibitem{Bako17}
S.~Bako, T.~Vogels, B.~McWilliams, M.~Meyer, J.~Nov\'ak, A.~Harvill, P.~Sen,
  T.~DeRose, and F.~Rousselle.
\newblock Kernel-predicting convolutional networks for denoising monte carlo
  renderings.
\newblock {\em ACM Transactions on Graphics (TOG) (Proceedings of SIGGRAPH
  2017)}, 36(4), July 2017.

\bibitem{balduzzi2016neural}
D.~Balduzzi, B.~McWilliams, and T.~Butler-Yeoman.
\newblock Neural taylor approximations: Convergence and exploration in
  rectifier networks.
\newblock {\em arXiv preprint arXiv:1611.02345}, 2016.

\bibitem{bauszat2015general}
P.~Bauszat, M.~Eisemann, E.~Eisemann, and M.~Magnor.
\newblock General and robust error estimation and reconstruction for monte
  carlo rendering.
\newblock In {\em Computer Graphics Forum}, volume~34, pages 597--608. Wiley
  Online Library, 2015.

\bibitem{bauszat2011guided}
P.~Bauszat, M.~Eisemann, and M.~Magnor.
\newblock Guided image filtering for interactive high-quality global
  illumination.
\newblock In {\em Computer Graphics Forum}, volume~30, pages 1361--1368. Wiley
  Online Library, 2011.

\bibitem{resources16}
B.~Bitterli.
\newblock Rendering resources, 2016.
\newblock https://benedikt-bitterli.me/resources/.

\bibitem{bitterli16nonlinearly}
B.~Bitterli, F.~Rousselle, B.~Moon, J.~A. Iglesias-Guiti\'an, D.~Adler,
  K.~Mitchell, W.~Jarosz, and J.~Nov\'ak.
\newblock Nonlinearly weighted first-order regression for denoising monte carlo
  renderings.
\newblock {\em Computer Graphics Forum (Proceedings of EGSR)}, 35(4), June
  2016.

\bibitem{bonneel2015blind}
N.~Bonneel, J.~Tompkin, K.~Sunkavalli, D.~Sun, S.~Paris, and H.~Pfister.
\newblock Blind video temporal consistency.
\newblock {\em ACM Transactions on Graphics (TOG)}, 34(6):196, 2015.

\bibitem{Boughida2017Bayesian}
M.~Boughida and T.~Boubekeur.
\newblock Bayesian collaborative denoising for monte carlo rendering.
\newblock {\em Computer Graphics Forum}, 36(4):137--153, 2017.

\bibitem{chaitanya2017interactive}
C.~R.~A. Chaitanya, A.~S. Kaplanyan, C.~Schied, M.~Salvi, A.~Lefohn,
  D.~Nowrouzezahrai, and T.~Aila.
\newblock Interactive reconstruction of monte carlo image sequences using a
  recurrent denoising autoencoder.
\newblock {\em ACM Transactions on Graphics (TOG)}, 36(4):98, 2017.

\bibitem{cook1984distributed}
R.~L. Cook, T.~Porter, and L.~Carpenter.
\newblock Distributed ray tracing.
\newblock In {\em ACM SIGGRAPH Computer Graphics}, volume~18, pages 137--145.
  ACM, 1984.

\bibitem{eilertsen2017hdr}
G.~Eilertsen, J.~Kronander, G.~Denes, R.~K. Mantiuk, and J.~Unger.
\newblock Hdr image reconstruction from a single exposure using deep cnns.
\newblock {\em ACM Transactions on Graphics (TOG)}, 36(6):178, 2017.

\bibitem{gharbi2016deep}
M.~Gharbi, G.~Chaurasia, S.~Paris, and F.~Durand.
\newblock Deep joint demosaicking and denoising.
\newblock {\em ACM Transactions on Graphics (TOG)}, 35(6):191, 2016.

\bibitem{glorot2010understanding}
X.~Glorot and Y.~Bengio.
\newblock Understanding the difficulty of training deep feedforward neural
  networks.
\newblock In {\em Proceedings of the Thirteenth International Conference on
  Artificial Intelligence and Statistics}, pages 249--256, 2010.

\bibitem{he2016deep}
K.~He, X.~Zhang, S.~Ren, and J.~Sun.
\newblock Deep residual learning for image recognition.
\newblock In {\em Proceedings of the IEEE Conference on Computer Vision and
  Pattern Recognition}, pages 770--778, 2016.

\bibitem{hinton2006reducing}
G.~E. Hinton and R.~R. Salakhutdinov.
\newblock Reducing the dimensionality of data with neural networks.
\newblock {\em science}, 313(5786):504--507, 2006.

\bibitem{jarosz2012theory}
W.~Jarosz, V.~Sch{\"o}nefeld, L.~Kobbelt, and H.~W. Jensen.
\newblock Theory, analysis and applications of 2d global illumination.
\newblock {\em ACM Transactions on Graphics (TOG)}, 31(5):125, 2012.

\bibitem{jensen2003monte}
H.~W. Jensen, J.~Arvo, P.~Dutre, A.~Keller, A.~Owen, M.~Pharr, and P.~Shirley.
\newblock Monte carlo ray tracing.
\newblock In {\em ACM SIGGRAPH}, pages 27--31, 2003.

\bibitem{LBF}
N.~K. Kalantari, S.~Bako, and P.~Sen.
\newblock {A Machine Learning Approach for Filtering Monte Carlo Noise}.
\newblock {\em ACM Transactions on Graphics (TOG) (Proceedings of SIGGRAPH
  2015)}, 34(4), 2015.

\bibitem{Kingma2014Adam}
D.~P. Kingma and J.~Ba.
\newblock Adam: A method for stochastic optimization.
\newblock {\em Computer Science}, 2014.

\bibitem{lai2018learning}
W.-S. Lai, J.-B. Huang, O.~Wang, E.~Shechtman, E.~Yumer, and M.-H. Yang.
\newblock Learning blind video temporal consistency.
\newblock In {\em Proceedings of the European Conference on Computer Vision
  (ECCV)}, pages 170--185, 2018.

\bibitem{lang2012practical}
M.~Lang, O.~Wang, T.~Aydin, A.~Smolic, and M.~Gross.
\newblock Practical temporal consistency for image-based graphics applications.
\newblock {\em ACM Transactions on Graphics}, 31(4):34, 2012.

\bibitem{mao2016image}
X.~Mao, C.~Shen, and Y.-B. Yang.
\newblock Image restoration using very deep convolutional encoder-decoder
  networks with symmetric skip connections.
\newblock In {\em Advances in Neural Information Processing Systems}, pages
  2802--2810, 2016.

\bibitem{moon2014adaptive}
B.~Moon, N.~Carr, and S.-E. Yoon.
\newblock Adaptive rendering based on weighted local regression.
\newblock {\em ACM Transactions on Graphics (TOG)}, 33(5):170, 2014.

\bibitem{pharr2016physically}
M.~Pharr, W.~Jakob, and G.~Humphreys.
\newblock {\em Physically based rendering: From theory to implementation}.
\newblock Morgan Kaufmann, 2016.

\bibitem{ramamoorthi2007first}
R.~Ramamoorthi, D.~Mahajan, and P.~Belhumeur.
\newblock A first-order analysis of lighting, shading, and shadows.
\newblock {\em ACM Transactions on Graphics (TOG)}, 26(1):2, 2007.

\bibitem{rousselle2011adaptive}
F.~Rousselle, C.~Knaus, and M.~Zwicker.
\newblock Adaptive sampling and reconstruction using greedy error minimization.
\newblock In {\em ACM Transactions on Graphics (TOG)}, volume~30, page 159.
  ACM, 2011.

\bibitem{rousselle2013dfc}
F.~Rousselle, M.~Manzi, and M.~Zwicker.
\newblock Robust denoising using feature and color information.
\newblock {\em Computer Graphics Forum}, 32(7):121--130, 2013.

\bibitem{wang2004image}
Z.~Wang, A.~C. Bovik, H.~R. Sheikh, and E.~P. Simoncelli.
\newblock Image quality assessment: from error visibility to structural
  similarity.
\newblock {\em IEEE transactions on image processing}, 13(4):600--612, 2004.

\bibitem{FRCNN}
X.~Yang, D.~Wang, W.~Hu, L.~Zhao, X.~Piao, D.~Zhou, Q.~Zhang, B.~Yin, Q.~Cai,
  and X.~Wei.
\newblock Fast reconstruction for monte carlo rendering using deep
  convolutional networks.
\newblock {\em IEEE Access}, 7:21177--21187, 2019.

\bibitem{zhang2017beyond}
K.~Zhang, W.~Zuo, Y.~Chen, D.~Meng, and L.~Zhang.
\newblock Beyond a gaussian denoiser: Residual learning of deep cnn for image
  denoising.
\newblock {\em IEEE Transactions on Image Processing}, 2017.

\bibitem{zwicker2015recent}
M.~Zwicker, W.~Jarosz, J.~Lehtinen, B.~Moon, R.~Ramamoorthi, F.~Rousselle,
  P.~Sen, C.~Soler, and S.-E. Yoon.
\newblock Recent advances in adaptive sampling and reconstruction for monte
  carlo rendering.
\newblock In {\em Computer graphics forum}, volume~34, pages 667--681. Wiley
  Online Library, 2015.

\end{thebibliography}

\end{document}